\documentstyle[12pt]{article}
\title{\huge Layering transitions of the spin§-1/2§ Ising model in a 
transverse magnetic field} 
\author{\bf  
 L. Bahmad, A. Benyoussef, A. Boubekri and H. Ez-Zahraouy 
\\
 Laboratoire de Magn\'{e}tisme et de la Physique
 des Hautes Energies
\\
Universit\'{e} Mohammed V, Facult\'{e} des Sciences, Avenue Ibn Batouta,  B.P. 1014
\\Rabat, Morocco
}
\date{ }
\begin{document}
\maketitle 
\vspace{3cm} 
\begin{abstract}
\mbox{  } Using mean field theory, the effect of the transverse magnetic field on the layering and wetting transitions of the spin-$1/2$ Ising model with longitudinal magnetic field is studied. At a fixed value of the temperature smaller than the wetting temperature, such system exibits a sequence of layering transitions above a critical value of the transverse magnetic field, wich depends on the temperature and surface magnetic field.

\end{abstract}
%\vskip 4cm
\noindent
%----------------------------------- \\ 
%PACS :75.10.Jm; 75.30.Ds

\newpage

\section{Introduction}
\mbox{  } Several authors have studied the wetting of spin$-1/2$ Ising model in a 
transverse magnetic field. Much attention has been paid to the properties of layered
structures consisting of alternating magnetic materials. The most commonly
studied magnetic multilayers are those of ferromagnetic transition metal such
as Co or Ni. Many experiments have shown that the magnetization enhancement
exists in multilayered films consisting of magnetic layers. It was found that
ferromagnetic coupling can exist between magnetic layers. From the
theoretical point of view, great interest has been paid to spin wave
excitations as well as critical phenomena. The study of thin films is partly motivated by the development
of new growth and characterization techniques, but perhaps more so by the
discovery of many exciting new properties, some quite unanticipated. These
include, more recently, the discovery of enormous values of magnetoresistance
in magnetic multilayers far exceeding those found in single layer films and
the discovery of oscillatory interlayer coupling in transition metal multilayers.
These experimental studies have motivated much theoritical work. However
these developments are to a large extent powered by $<<$ materials engineering $>>$
and the ability to control and understand the growth of thin layers only a
few atoms thick. Multilayer films adsorbed on attractive substrates may
exibit a variety of possible phase transitions, as has been
reviewed by Pandit {\it et al.}, and Ebner {\it et al.}\\
\mbox{  }This paper is organized as follows. Section 2 describes the model
and the method. In section 3 we present results and discussions.
\newpage
\section{Model and method}
We consider N coupled ferromagnetic square layers in a transverse and longitudinal magnetic fields. The Hamiltonian of the system is given by
\begin{equation}  
§H=-\sum_{<i,j>}J_{ij}S_{i}^{z}S_{j}^{z}-\sum_{i}(\Omega S_{i}^{x}+H_{i}S_{i}^{z})§
\end{equation}
where, \( S_{i}^{\alpha} ,(\alpha=x,z) \) are the Pauli matrices, $\Omega$ is the transverse magnetic field in the x direction and \(H_{i}\) is the magnetic field in the z direction, defined by:
\begin{equation}   
H_{i}=\left\{ \begin{array}{lll}
	      H+H_{s1} & \mbox{for}&  i=1  \\
	      H        & \mbox{for}&  1 < i < N \\
	      H+H_{s2} & \mbox{for}&  i=N 
	      \end{array}
	\right.
\end{equation}
The surface magnetic field $H_{s1}$ is applied on the first layer $(i=1)$, the last layer $(i=N)$ is under a surface magnetic field $H_{s2}$ and H is the bulk magnetic field.\\
We introduce the effective field 
\begin{equation}  
§H_{0}^{eff}=-\sum_{i}H_{0}^{i}=-\sum_{i}(k_{i}S_{i}^{z}+\Omega S_{i}^{x}) § 
\end{equation}
where
\begin{equation}
§k_{i}=\sum_{j}J_{ij}S_{j}^{z}+H_{i}§,
\end{equation}
 
and using the Pauli matrices one can write :
\begin{equation}
H_{0}^{i}=§\left(\begin{array}{cc}
	      §k_{i}§ & \Omega  \\
              \Omega        & §-k_{i}§                
	      \end{array}\right) §
\end{equation}
The eigen values of the Hamiltonian \(H_{i}^{0}\) are given by :\\
\begin{equation}
§\lambda_{i}^{\pm}=\pm\sqrt{k_{i}^{2}+\Omega^{2}}.§
\end{equation}
 
The magnetization per site can be written as 
\begin{equation}
§ m_{i}=\frac{TrS_{i}^{z}\exp(-\beta H_{0}^{eff})}{Tr\exp(-\beta H_{0}^{eff})}=[\frac{(k_{i}+\lambda_{i}^{+})^{2}-\Omega^{2}}{(k_{i}+\lambda_{i}^{+})^{2}+\Omega^{2}}]\tanh(\beta \sqrt{k_{i}^{2}+\Omega^{2}}).  §
\end{equation}
 
%Using the mean field theory, the free energy of the system can be %written as follows 
%\begin{equation}
%§ F=F_{0}+<H-H_{0}>_{0}.§
%\end{equation}
%By substitutions, we find 
%\begin{equation}
%§F_{0}=-\frac{1}{\beta}\log(Tr\exp(-\beta %H_{0}))=-\frac{1}{\beta}\sum_{i}\log(2\cosh(\beta\lambda_{i}^{+})§.
%\end{equation}
%Using the equations cited above the final form of the free energy is
%\begin{equation}
%§     F=-\frac{1}{\beta}\sum_{i}\log(2\cosh(\beta\lambda_{i}^{+}))+
%\frac{1}{2}\sum_{i}\sum_{j}J_{ij}m_{i}m_{j}§.
%\end{equation}
%For a simple cubic lattice, the magnetization and the free energy for %each plane are given, respectively, by the following expressions:
%\begin{equation}
%§m_{p}=\frac{1}{\lambda_{p}}[4m_{p}+r(m_{p+1}+m_{p-1}+H^{'})]\tanh(\be%ta^{'}\lambda_{p}) §
%\end{equation}
%and
%\begin{equation}
%§ F_{p}=-\frac{1}{\beta}\log(2\cosh(\beta %§\lambda_{p}))+\frac{1}{2}m_{p}(4m_{p}+r(m_{p+1}+m_{p-1})§ 
%\end{equation}
%with 
%\begin{equation}
%§\lambda_{p}=\sqrt{ (4m_{p}+r(m_{p+1}+m_{p-1})+H^{'})^2+\Omega^{'2}}, %§
%\end{equation}
%\\
%where, the reduced parameters are defined by:  
%$\beta^{'}=\beta J_{//}$ , $H^{'}=\frac{H}{J_{//}}$ ,  %$\Omega^{'}=\frac{\Omega}{J_{//}}$ and $r=\frac{J_{\perp}}{J_{//}}$.  % 
\newpage
\section{Results and discussion}
\subsection{Ground state in the absence of the transverse magnetic field}
\mbox{   }We focus our interest on a system corresponding to the case $J_{//}=J_{\perp}=J$.
The ground states, in absence of the transverse magnetic field, of the system are plotted in Figs. 1a, 1b. Indeed, the Fig. 1a
illustrates the ground state of the system with a surface magnetic field $H_{s1}/J$, applied on the first layer, whereas
the last layer is under a surface magnetic field $H_{s2}/J=-H_{s1}/J$.
The start point is a situation where all spins are down, $O^{N}$. For $H_{s1}/J\geq 1$, we have three transitions namely; the first transition, $O^{N}\leftrightarrow 1O^{N-1}$, arises for $H/J=1-H_{s1}/J$ , the second transition, $1O^{N-1}\leftrightarrow 1^{N-1}O$, occurs for $H/J=0$, while the last transition $1^{N-1}O\leftrightarrow 1^{N}$, occurs for $H/J=1+H_{s1}/J$. Thus, for $H_{s1}/J\leq 1$, we have only one transition, $O^{N}\leftrightarrow 1^{N}$. The notation $1^{p}O^{N-p}$ is a situation where p layers are spin-up while N-p layers are spin-down.
The Fig. 1b shows that the transitions arise for values of $H_{s1}/J$ above a critical value, $H^{c}_{s1}/J=N/(N-1)$. Indeed, for $H_{s1}/J < H^{c}_{s1}/J$, we find only one transition, $O^{N}\leftrightarrow 1^{N}$. While for $H_{s1}/J > H^{c}_{s1}/J$, the first transition, $O^{N}\leftrightarrow 1O^{N-1}$, exists for $H/J=1-H_{s1}/J$, and the last transition, $1^{N-1}O\leftrightarrow 1^{N}$, can be seen for $H/J=-1/(N-1)$.
  
\subsection{Phase diagrams}

\mbox{  }The system we are considering is formed with N ferromagnetic layers of a spin$-1/2$ Ising model with free bound conditions. In order to examine only the effect of the transverse magnetic field, $\Omega /J$, on the wetting and layering transitions, we fixe the temperature, T/J, at a value less than the wetting temperature, $T_{w}/J$. However, by increasing $\Omega /J$ we show the existence of a critical value, $\Omega _{w} /J$, above which a sequence of layering transitions occurs. This is called the wetting transverse magnetic field, which depends on the value of the temperature and the surface magnetic field, $H_{s1}/J$. On the other hand, when $T/J > T_{w}/J$, the layering transitions exist in absence of the transverse magnetic field, this means that the temperature effect is sufficient to produce layering transitions. \\   
%\mbox{  }We solve numerically the equations (11) and (12), in order to %estabilish the phase diagrams of the system, namely:
Fig. 2a shows the existence of a wetting temperature, $T_{w} /J$, above which the layering transitions occurs, while for $T/J \leq T_{w}/J$ there is only one layer transition. The phase diagram, which is displayed in Fig. 2b, shows that the layer transitions are found for $\Omega /J > \Omega_{w} /J$ with increasing transverse magnetic field and fixed temperature. The wetting transverse magnetic field $\Omega_{w} /J$, decreases when increasing temperatures, and it increases when decreasing surface magnetic field values, and vice versa .
Thus, the first and last layer transitions occcur in the absence of the transverse magnetic field, but only in the presence of the bulk magnetic field (Fig 2c). The other layering transitions arise for increasing transverse magnetic field.  
It is also interesting to examine the case corresponding to $H_{s2}/J=0$.
The phase diagrams in the plane $(H/J,\Omega /J)$ are plotted in figures 3a, 3b. The Fig. 3a is related to the situation: $H_{s1}/J< H^c_{s1}/J$ while the Fig. 3b corresponds to $H_{s1}/J> H^c_{s1}/J$. From Fig. 3a, we find that the layering transitions exist only for values of the transverse magnetic field greater than a critical value, which depends of the temperature and surface magnetic field.  Fig. 3b shows that the first layering transition arises although the absence of the transverse magnetic field, but only for increasing bulk magnetic field. The other layer transitions occur when increasing the transverse magnetic field, at fixed values of the temperature and surface magnetic field. Furthermore, the phase diagram in $(T/J,\Omega/J)$ plane is presented in Fig. 4. However, the wetting transverse magnetic field,$\Omega_{w}/J$, decreases with increasing temperature and/or decreasing surface magnetic field, and the wetting temperature, $T_{w}/J$, decreases with increasing transverse magnetic field and/or decreasing surface magnetic field. \\       
\mbox{~~~ }In conclusion, we have studied the effect of the transverse magnetic field on the layering and wetting transitions of a spin$-1/2$ Ising model in a longitudinal magnetic field. However, we have shown that the layering and wetting transitions appear above a critical transvere field, $\Omega_{w}/J$, which is a function of the temperature and surface magnetic field.

\newpage
\noindent{\bf References}
\begin{enumerate}

\item[[1]] A.Benyoussef and H. Ez-Zahraouy, Physica A,{\bf 206},196 (1994).  
\item[[2]] A. B. Harris, C. Micheletti and J. Yeomans, J. Stat. Phys. {\bf 84},323 (1996).

\item[[3]] A. Benyoussef and H. Ez-Zahraouy, J. Phys. {\it I} France {\bf 4}, 393 (1994).

\item[[4]] S. Dietrich and M. Schick, Phys. Rev B {\bf 31},4718 (1985).

\item[[5]] M. P. Nightingale, W. F. Saam and M. Schick, Phys. Rev. B {\bf 30},3830 (1984).

\item[[6]] R. Pandit, M. Schick and M. Wortis, Phys. Rev. B {\bf 26}, 8115 (1982).

\item[[7]] C. Ebner, C. Rottman and M. Wortis, Phys. Rev. B {\bf 28},4186  (1983).  

\item[[8]] C. Ebner and W. F. Saam, Phys. Rev. Lett. {\bf 58},587 (1987).

\item[[9]] C. Ebner and W. F. Saam, Phys. Rev. B {\bf 35},1822 (1987).

\item[[10]] C. Ebner, W. F. Saam and A. K. Sen, Phys. Rev. B {\bf 32},1558 (1987).

\item[[11]] S. Ramesh and J. D. Maynard, Phys. Rev. Lett. {\bf 49},47 (1982).

\item[[12]] S. Ramesh, Q. Zhang, G. Torso and J. D. Maynard, Phys. Rev. Lett. {\bf 52},2375 (1984).

\item[[13]] M. Sutton, S. G. J. Mochrie  and R. J. Birgeneou, Phys. Rev. Lett. {\bf 51},407 (1983);\\
S. G. J. Mochrie, M. Sutton, R. J. Birgeneou, D. E. Moncton and P. M. Horn, Phys. Rev. B {\bf 30},263 (1984).

\item[[14]] S. K. Stija, L. Passel, J. Eckart, W. Ellenson and H. Patterson, Phys. Rev. Lett. {\bf 51},411 (1983).

\item[[15]] D. A. Huse , Phys. Rev. B {\bf 30},1371 (1984).

\item[[16]] M. J. de Oliveira  and R. B. Griffiths , Surf. Sci. {\bf 71}, 687 (1978).

\item[[17]] C. Ebner and W. F. Saam, Phys. Rev. A {\bf 22},2776 (1980);\\
     ibid, Phys. Rev. A {\bf 23},1925 (1981);\\
     ibid, Phys. Rev. B {\bf 28},2890 (1983).

\item[[18]] S. J. Kennedy and S. J. Walker, Phys. Rev. B {\bf 30},1498 (1984).

\item[[19]] P. Wagner and K. Binder, Surf. Sci. {\bf 175},421 (1986).

\item[[20]] K. Binder and D. P. Landau, Phys. Rev. B {\bf 37}, 1745 (1988). 

\item[[21]] A. Patrykiejew A., D. P. Landau and K. Binder, Surf. Sci. {\bf 238}, 317 (1990).

\item[[22]] H. Nakanishi and M. E. Fisher, J. Chem. Phys. {\bf 78},3279  (1983).

\item[[23]] E. Bruno, U. Marini, B. Marconi and R. Evans, Physica A {\bf 141A}, 187 (1987).

\item[[24]] R. Kariotis  and J. J. Prentis, J. Phys. A{\bf 19}, L 455 (1986).

\item[[25]] R. Kariotis, B. Yang and H. Suhl, Surf. Sci. {\bf 173},283 (1986).

\item[[26]] J. M. Luck, S. Leibter and B. Derrida, J. Phys. France {\bf 44}, 1135 (1983).

\item[[27]] K. K. Mon and W. F. Saam, Phys. Rev. B {\bf 23}, 5824 (1981).

\item[[28]] W. F. Saam, Surf. Sci. {\bf 125},253 (1983).

\item[[29]] L. P. Kadanoff, Ann. Phys. (NY) {\bf 100}, 359 (1976).

\item[[30]] M. P. Nightingale, J. Appl. Phys. {\bf 53}, 7927 (1982).

\item[[31]] H. J. Hermann, J. Phys. Lett. {\bf 100A}, 256 (1984).

\item[[32]] M. P. Nightingale, Physica {\bf 83A}, 561 (1979).

\item[[33]] R. Pandit and M. Wortis, Phys. Rev. B {\bf 25}, 3226 (1982).  

\item[[34]] K. Binder and D. P. Landau, J. Appl. Phys. {\bf 57}, 3306 (1985).
\end{enumerate}
\newpage
\noindent{\bf Figure Captions}\\
\\
\noindent{\bf Figure 1.}: Phase diagrams for $T=0$ and $\Omega =0$ in the $(H/J,H_{s1}/J)$ plane (N=20 layers)\\
      a) $H_{s2}/J=-H_{s1}/J$ \\
      b) $H_{s2}/J=0$

\noindent{\bf Figure 2.}: Phase diagrams for $H_{s2}/J=-H_{s1}/J$ and $N=20$. \\
      a)$H_{s1}/J=0.8$,$\Omega /J=0$ in the $(H/J,T/J)$ plane. \\
      b)$H_{s1}/J=0.8$,$T/J=2.0$ in the $(H/J,\Omega /J)$ plane.\\        c)$H_{s1}/J=1.2$,$T/J=2.0$ in the $(H/J,\Omega /J)$ plane.
	  
\noindent{\bf Figure 3.}: Phase diagrams for $H_{s2}/J=0$, $T/J=2.0$ and $N=20$ in the $(H/J,\Omega/J)$ plane.\\
	a)$H_{s1}/J=0.9$    \\    
        b)$H_{s1}/J=1.6$ 
	  
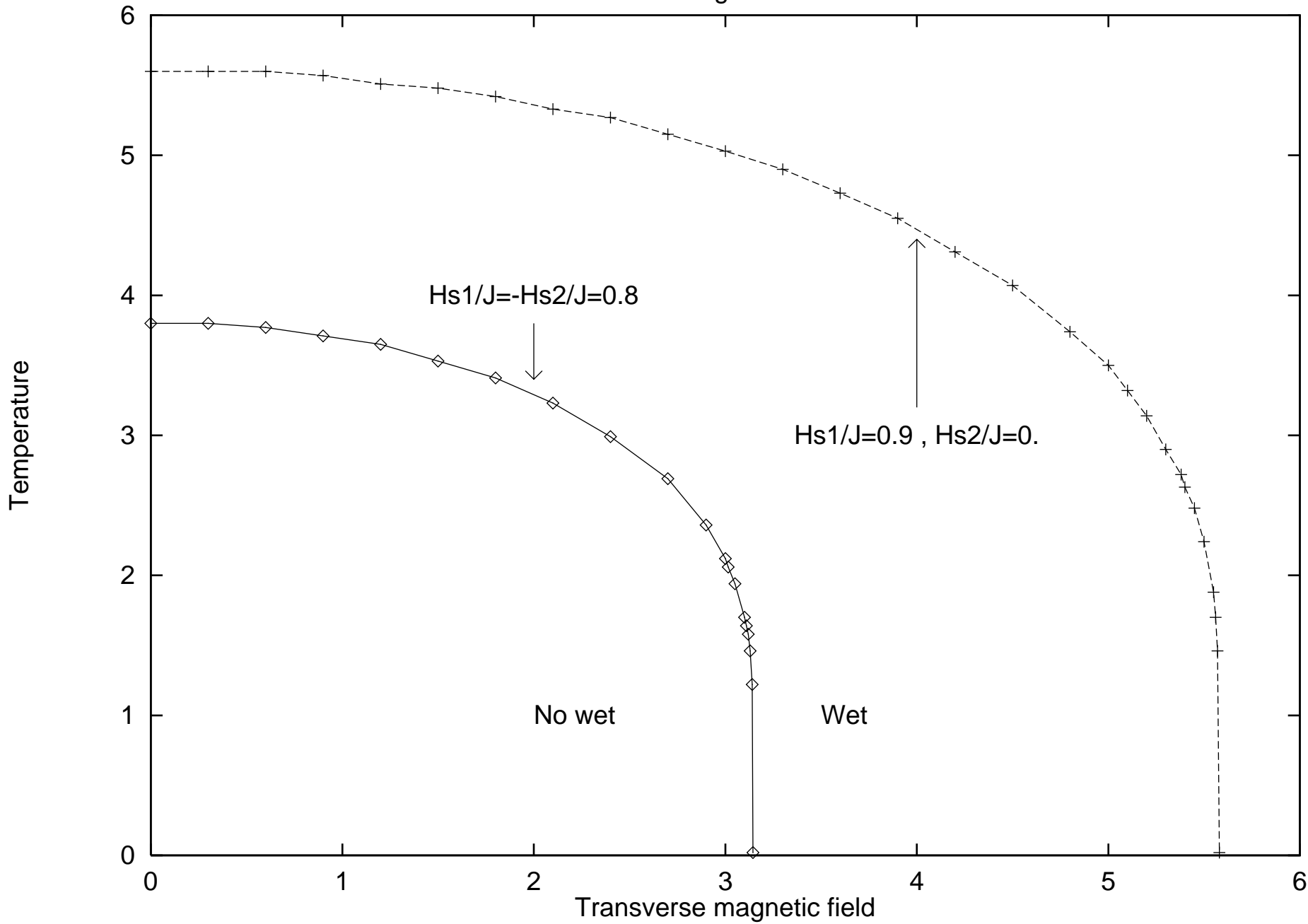
\noindent{\bf Figure 4.}: Phase diagram in $(T/J,\Omega/J)$ plane,
for both cases: $(H_{s1}/J=0.8,H_{s2}/J=-H_{s1}/J)$ and $(H_{s1}/J=0.9,H_{s2}/J=0.)$, and $N=20$. 

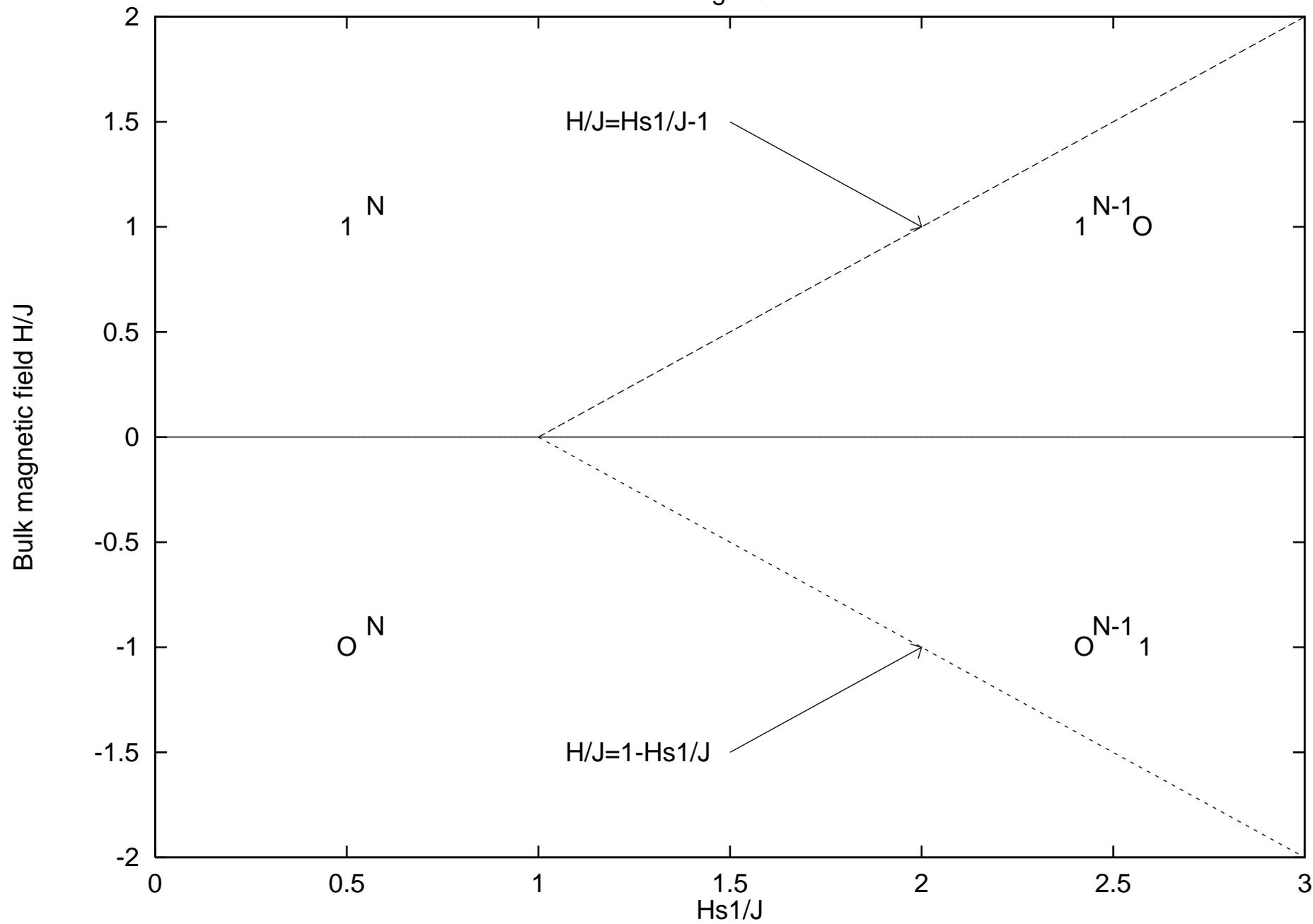
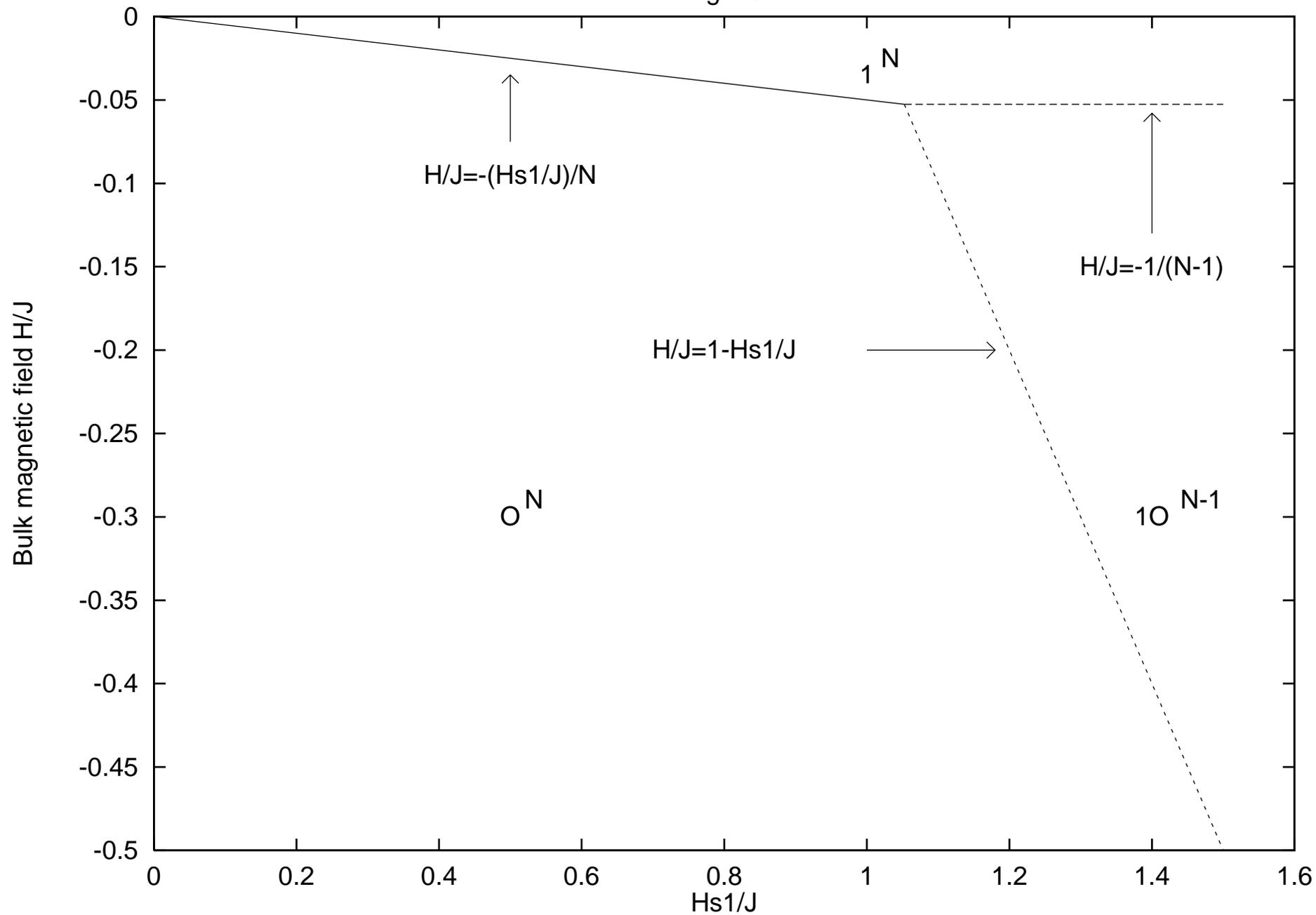
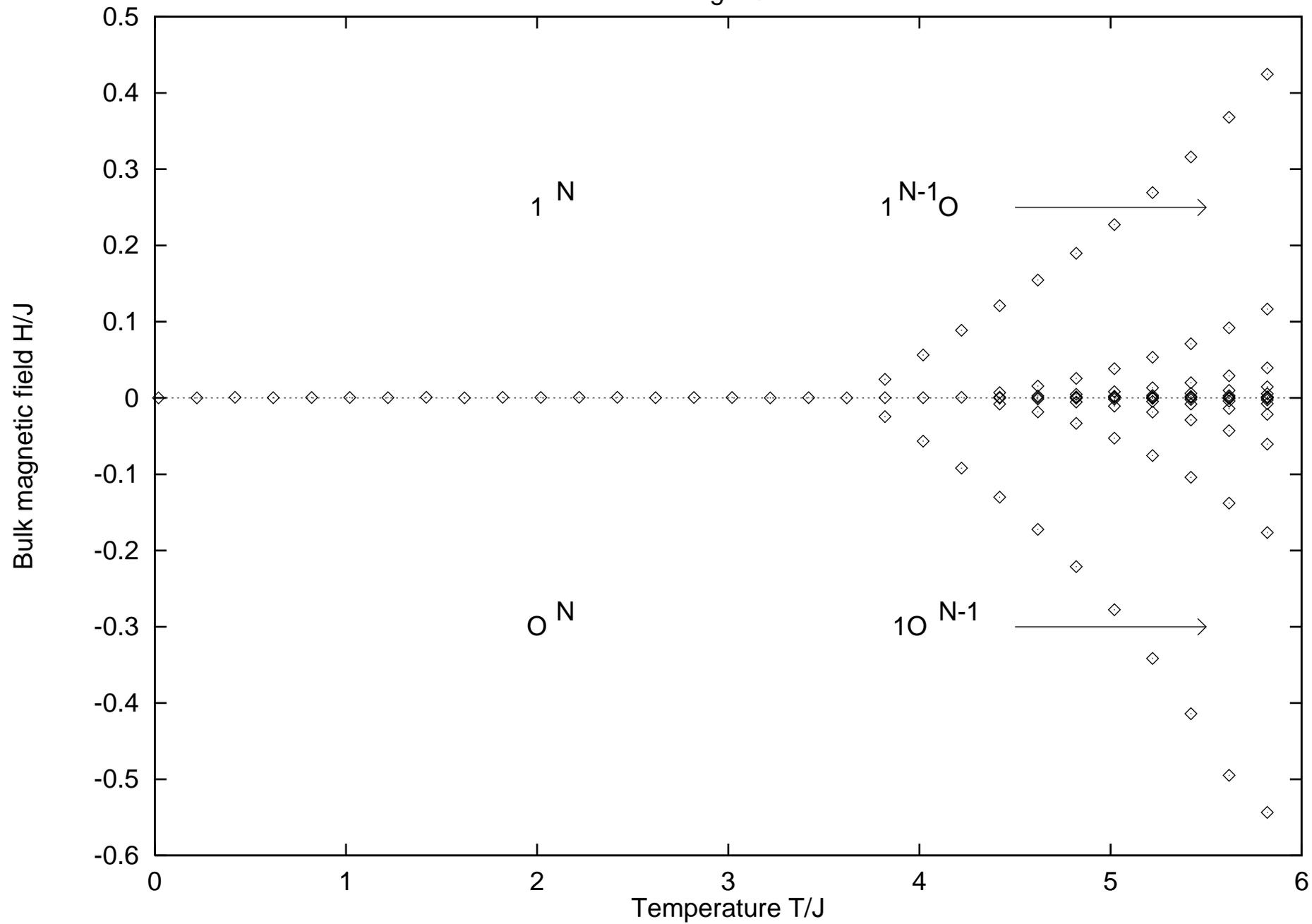
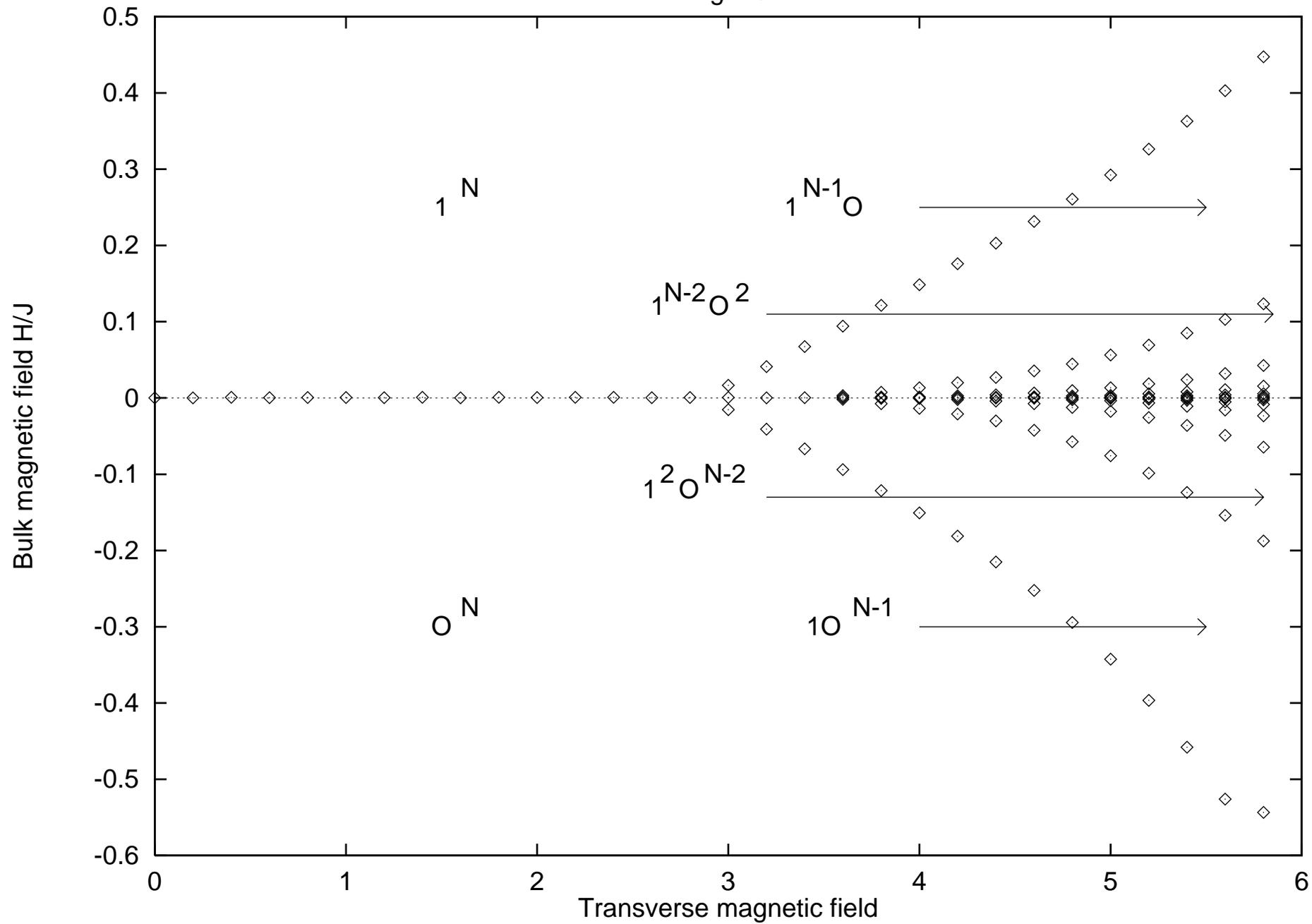
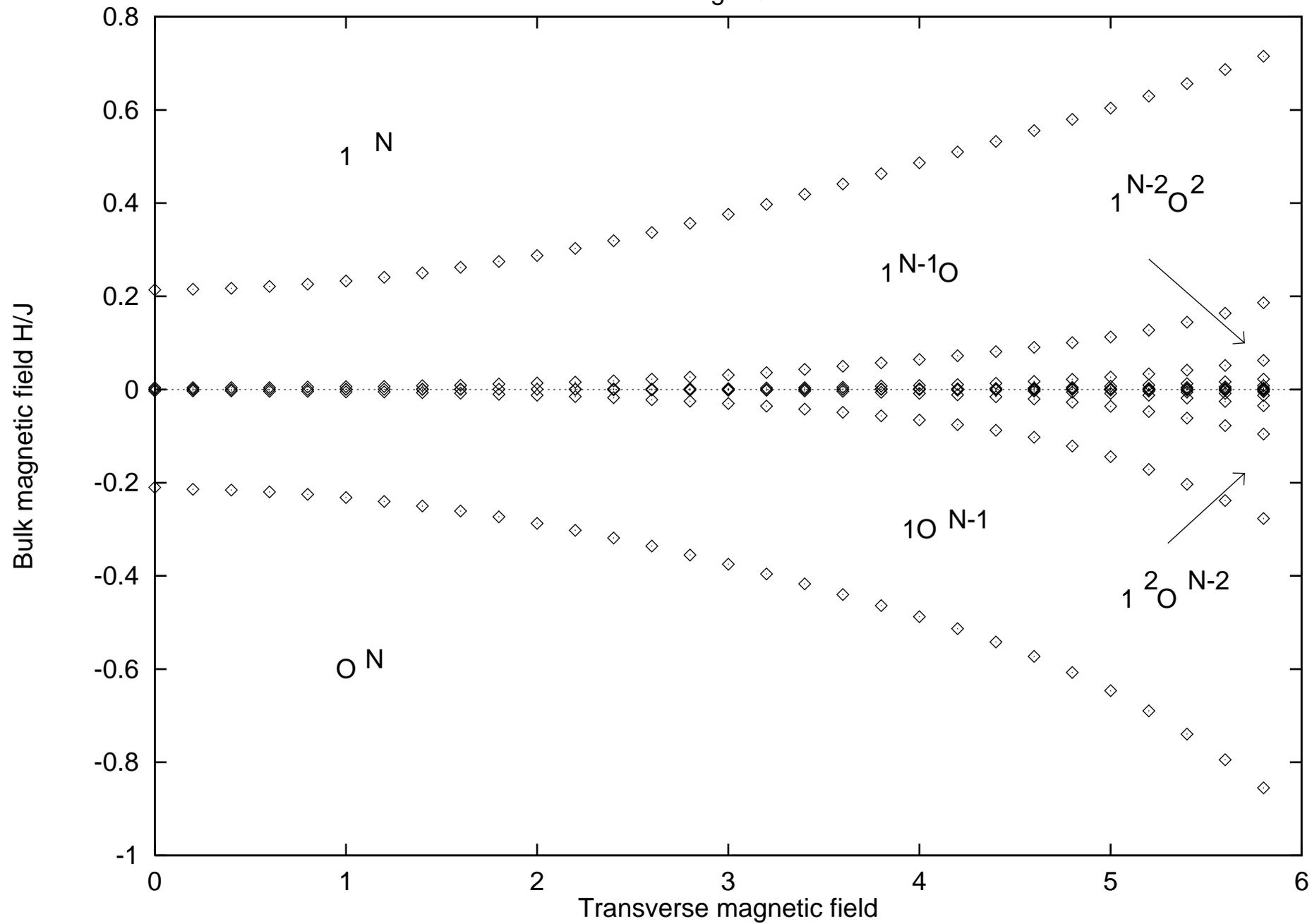
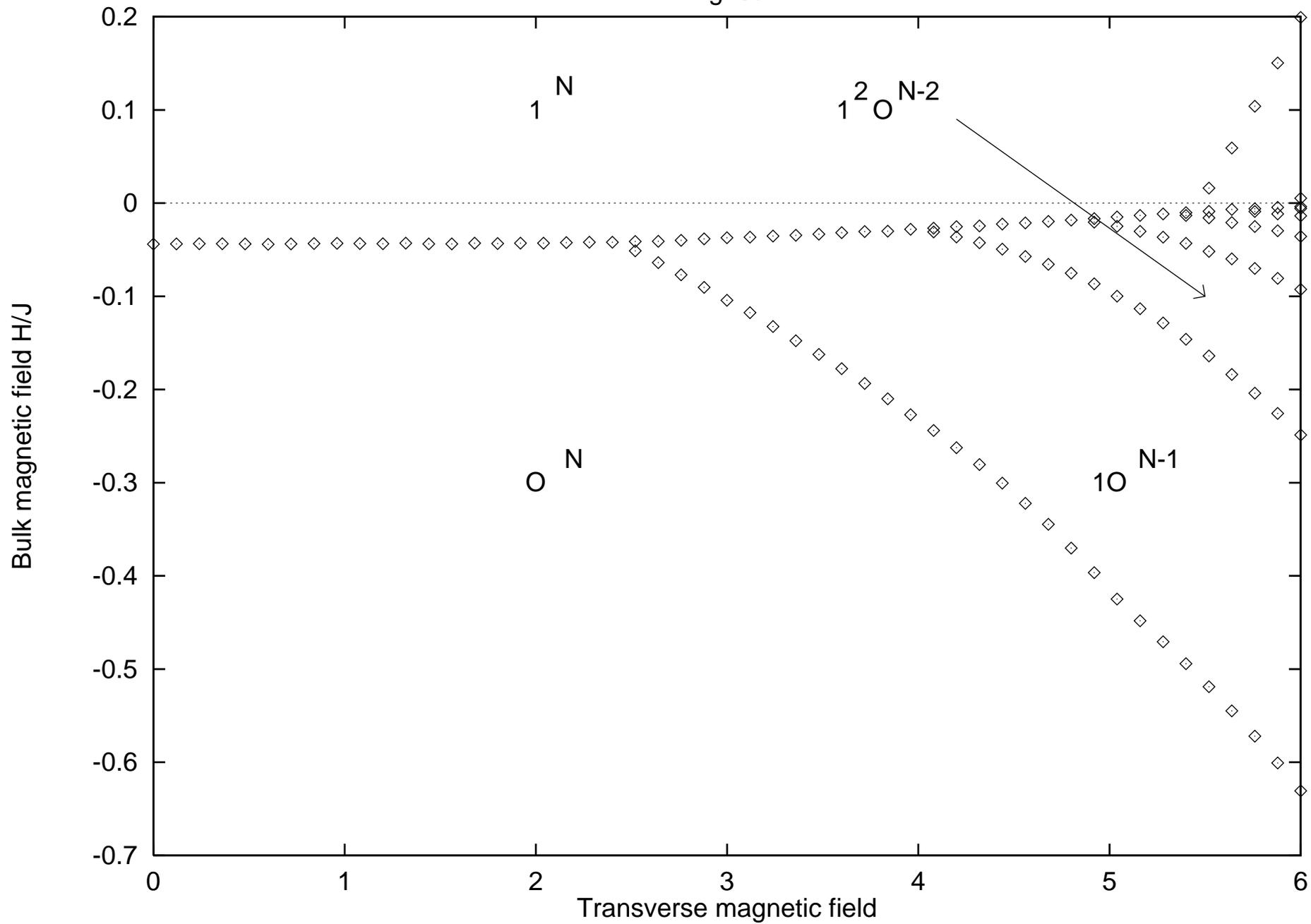
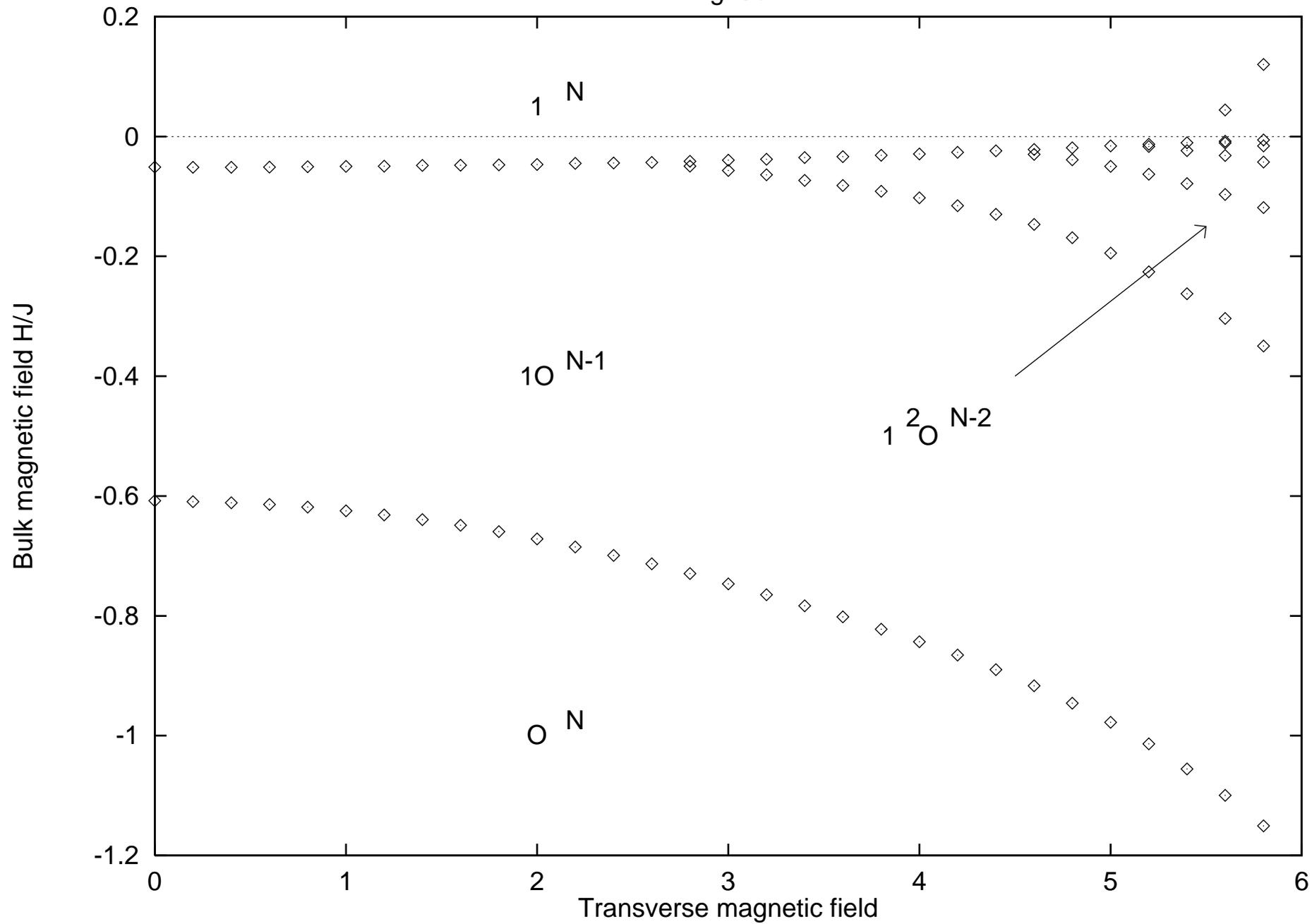
\end{document}